\documentclass[11pt,reqno]{amsart}
\usepackage{amsaddr}
\usepackage[table]{xcolor}

\usepackage[margin=2.7cm]{geometry}
\usepackage{array}
\usepackage{booktabs}
\usepackage[dvipdfmx]{graphicx}
\usepackage[sectionbib]{natbib} 
\usepackage{amsfonts}
\usepackage{amssymb,amsthm}
\usepackage{amsmath}
\usepackage{float}
\usepackage[font=small,labelfont=bf]{caption}

\theoremstyle{definition}

\theoremstyle{remark}

\numberwithin{equation}{section}

\begin{document}

\title{Macroeconomic factors for inflation in Argentine 2013-2019}

\author{Manuel Lopez Galvan}

\address{Facultad de Ciencias Exactas y Naturales - Universidad de Buenos Aires  }

\email{amlopezgalvan@gmail.com , mlopezgalvan@hotmail.com }
\thanks{}


\date{}

\dedicatory{}

\begin{abstract}
The aim of this paper is to investigate the use of the Factor Analysis in order to identify the role of the relevant macroeconomic variables in driving the inflation. The Macroeconomic predictors that usually affect the inflation are summarized using a small number of factors constructed by the principal components. This allows us to identify the crucial role of money growth, inflation expectation and exchange rate in driving the inflation. Then we use this factors to build econometric models to forecast inflation. Specifically, we use univariate and multivariate models such as classical autoregressive, Factor models and FAVAR models. Results of forecasting suggest that models which incorporate more economic information outperform the benchmark. Furthermore, causality test and impulse response are performed in order to examine the short-run dynamics of inflation to shocks in the principal factors.    

\end{abstract}

\maketitle
\section{Introduction}
In the last 10 years, Argentine has experienced one of the highest inflation rates of any country in the world and therefore it is an interesting case to study the effect and the relationships of their main macroeconomics variables on the inflation rate. There are various works on the effects of the relationship of monetary policy and inflation and also measures as the output gap or interest rate. Basco, D 'Amato and \citet{Garegnani(2006)} studied the short-run dynamics of money and prices under high and low inflation and found that that proportionality holds for high inflation period but weakness under low regime. 
 Most economists agree that high inflation usually begins when the Central Bank issues money to finance a fiscal disequilibrium; this argument was initially developed by \citet{Cagan(1956)} to explain hyperinflation. The problem of inflation has been overcome decades ago by developed countries focusing their attention primarily on monetary tools. In Argentine, the exchange rate plays a very import role on tradable goods pricing, however, it has been seen that the inflation has continued for long term without suddenly changes on exchange rate. Wages are a very important component of production costs and impact on non-tradable goods, furthermore, firms and agents take into account the expected rate of inflation on pricing decisions and wage contracts. All of these reasons may have effect on inflation, however, it is not clear why a high rate persists, although the deficit does not grow, although there are no sharp variations in the exchange rate. Thus, it is possible that prices dynamics could also be explained by unobserved or latent factors related to the observable pressures of inflation and therefore it should be studied from a multivariate point of view by using Factor Analysis. In this work we investigate what are the most important determinant of actual inflation and we will use this information to build forecast models.  

Classical econometrics models, such as AR, ADL or VAR can be used for simultaneously modelling the interaction of only a handful of variables, for example, in VAR models the number of parameters to be estimated increases geometrically with the number of variables and proportionally with the number of lags included. Factor Analysis is a technique used to discover clusters of variables so that the variables of each group are highly correlated, in this way this mode reduces a number of variables intercorrelated to a lower number of factors, which explain most of variability of each of the variables. \citet*{Stock and Watson(2002)} had explored the use of Factor Analysis into forecast models. Given a large number of macroeconomic variables, a time series of the factors are estimated from the predictors and then a linear regression between the variable to be forecast and the factors is performed.    

The structure of this paper is as follows: Section 2 we develop the theoretical framework for factor models with their respective assumptions, in Section 3 we analyze the data to be used and we verify the assumption of the framework, in Section 4 we develop the Factor Analysis and we analyze the relation between Factors and variables. In section 5 we start studying ceteris paribus effect on static factor models, then Granger-Causality test and impulse response analysis and finally we perform forecasting through Factor and FAVAR models. Section 6 concludes.

\section{Empirical Factor Model}\label{theoFactormodel}  
\subsection{Forecast Model}
The forecasting model to be applied to inflation in Argentine follows the two step Principal Component approach of  \citet*{Stock and Watson(2002)}. First, a time series of the factors are estimated from the predictors by using Principal Components methodology and then a linear regression is performed between the variable to be forecast and the estimated factors.
More precisely, recalling \citet*{Stock and Watson(2002)},  let $\pi_t$ be the time series to be forecast and let $X_t=(X_{ti})$ be a $N$-dimensional multiple time series of predictors  $(t=1,2,...,T , \ i=1,2,..,N)$, the factor model representation for the data $(X_{ti},\pi_t)$ is,

\begin{eqnarray}
X_t= F_t \Lambda^{\prime}+ \epsilon_t \label{FactorModel}
\end{eqnarray}
and
\begin{eqnarray}
\pi_{t+h}=\eta + \alpha^{\prime}F_t +\beta^{\prime} Z_t + \varepsilon_{t+h} \label{FactorModelForecast}
\end{eqnarray}
\bigskip

where $F_t \in T\times r$ matrix of $r$ factors, $\Lambda \in N \times r$ is the factor loadings matrix, $\epsilon_t$ is the matrix of idiosyncratic error, cross-sectionally independent and temporally iid, where $Z_t$ is a $m\times 1$ vector of observed variables (e.g. lags of $\pi_t$) and $\varepsilon_{t+h}$ is the forecast error.     

\bigskip

In order to estimate equation \ref{FactorModel} one needs to estimate $F_t$ and the matrix $\Lambda$, assuming the restrictions $\Lambda^{\prime}\Lambda=I$ and $T>N$ we minimize the following function,

\begin{eqnarray}
V(F,\Lambda)=\sum^N_{i=1}\sum^T_{t=1} (X_{it}-\lambda_iF_t^{\prime})^2 = tr \ (X^{\prime}-\Lambda F^{\prime})^{\prime}(X^{\prime}-\Lambda F^{\prime}) \label{ObjfunFactorModel}
\end{eqnarray}

\bigskip

where $\lambda_i$ is the ith row of $\Lambda$ and $tr$ denotes the usual trace.  
Minimizing equation \ref{ObjfunFactorModel} is equivalent to maximizing $tr \ (\Lambda^{\prime} X^{\prime}X \Lambda)$ and this is equivalent to the problem of finding maximun variance linear combinations of components of vector $X\Lambda$ under $\Lambda^{\prime}\Lambda=I$. Therefore it is the classical principal component problem and the resulting principal components estimator of $F$ is then,

\begin{eqnarray}
\tilde{F}=X\tilde{\Lambda}  
\end{eqnarray}
\bigskip

where $\tilde{\Lambda}$ has as its columns the first few eigenvectors of the covariance matrix of $X$.

\subsubsection{Out of the sample Static forecast (Or, one step ahead)}

Let $P<T$ be a number of observations, to build static out of sample forecast for $\pi_{t}$ on $\lbrace t=T-P+1,...,T \rbrace $    
, firstly we form principal components of the data ${\lbrace X_t \rbrace}^{T} _{t=1}$ to serve as estimates of the factors and then up to moment $T-P$ we perform linear regression of $\pi_t$ on factor estimators $\tilde{F}_{t-1}$ and $Z_{t-1}$. Let $\hat{\alpha},\hat{\beta}$ and $\hat{\eta}$ be the estimated coefficient then the out of the sample forecasts is constructed as, 

\begin{eqnarray}\label{static_forecast}
\widehat{\pi}_{t}= \hat{\eta} + \hat{\alpha}^{\prime}\tilde{F}_{t-1} + \hat{\beta}^{\prime}Z_{t-1}            
\end{eqnarray} 
\bigskip  

where $t$ ranges from $T-P+1$ up to $T$.   

\subsubsection{Out of the sample Dynamic forecast}

To construct Dynamic forecast under Factor Model methodology we would need to assume a joint dynamics for $(F_t,\pi_t)$, then two equations are estimated and the joint dynamics are given by, 

\begin{eqnarray}\label{General_FAVAR}
\begin{bmatrix}
F_t\\
\pi_t
\end{bmatrix}
=\phi(L)
\begin{bmatrix}
F_{t-1}\\
\pi_{t-1}
\end{bmatrix}
+ \varepsilon_t
\end{eqnarray}

\bigskip

where $\phi(L)$ is a conformable lag polynomial, Equation \ref{General_FAVAR} is referred to as a FAVAR model as formal introduced by \citet{Bernanke(2005)}. More precisely, the FAVAR model assumes that $F_t$ and $\pi_t$ jointly follows a VAR process. Then, the forecast is obtained recursively by using the factor estimators $\tilde{F_t}$,    
 
\begin{eqnarray}\label{dynamic_forecast1}
\widehat{\tilde{F}}_{t}=  \widehat{\phi_{11}} \widehat{\tilde{F}}_{t-1} +  \widehat{\phi_{12}}\widehat{\pi}_{t-1} \\
\widehat{\pi}_{t}= \widehat{\phi_{21}}\widehat{\tilde{F}}_{t-1} + \widehat{\phi_{22}} \widehat{\pi}_{t-1}   \label{dynamic_forecast2}       
\end{eqnarray} 
\bigskip  
    
where $t$ ranges from $T-P+1$ up to $T$ and $\widehat{\phi_{ij}}$ are the least square estimators up to moment $T-P$.

\subsection{Forecast evaluation}
\subsubsection{Performance Measures}
In order to compare the forecast performance two metrics will be used in this work; the Root Mean Square Error (RMSE) and the Theil’s U statistics. Autoregressive model (AR(1)) will be used as benchmark. Given a number of observations $P<T$, the Root Mean Square Error (RMSE) is defined as,    

\bigskip

\begin{eqnarray}
RMSE_{P}= \left[ \dfrac{1}{P}\sum^T_{t=T-P+1} (\pi_{t} - \hat{\pi}_{t})^2 \right]^{1/2}
\end{eqnarray} 

\bigskip

where $\lbrace \pi_t\rbrace$ are the observed values and $\lbrace \hat{\pi}_t\rbrace$ are the forecast values.
Theil’s U statistics or Theil’s coefficient of inequality is a relative Root Mean Square Error, more specifically we define the U-Theil as, 

\bigskip

\begin{eqnarray}
U_{P}= \dfrac{ \left[ \dfrac{1}{P}\sum^T_{t=T-P+1} (\pi_{t} - \hat{\pi}_{t})^2 \right]^{1/2}}{ \left[ {\frac{1}{P} \sum^T_{t=T-P+1} \pi_{t}^2} \right]^{1/2}  + \left[ {\frac{1}{P} \sum^T_{t=T-P+1} \hat{\pi}_{t}^2 } \right]^{1/2}}
\end{eqnarray} 

\bigskip

The U-Theil measure is bounded between 0 and 1 and values closer to 0 indicate better forecasting performance of the evaluated models.
Other statistics are given as ratios to RMSE of the autoregressive (AR) model and U-Theil. More precisely we define, 

\bigskip

\begin{eqnarray}
RRMSE_P=\frac{RMSE_P(Model)}{RMSE_P(AR)}
\  \ ,  \ RU_P=\frac{U_P(Model)}{U_P(AR)}
\end{eqnarray} 

Therefore, when the ratio is less than unity, the model (in RMSE or U-Theil - context) is better than the autoregresive benchmark model.

\bigskip

\subsubsection{Tests to compare forecast power}\label{DM}
\citet{Diebold and Mariano(1995)} developed a hypothesis test to compare forecast power of two competing forecasts. Given two competing forecast $\widehat{\pi}_{1t}$ and $\widehat{\pi}_{2t}$ of a particular actual $\pi_t$, let $\lbrace e_{1t} \rbrace$ and $\lbrace e_{2t} \rbrace$ be the associated forecast errors and let $g(e_{it})$ be the loss function of the forecast error. The loss differential is defined as $d_t=g(e_{1t})-g(e_{2t})$, the null hypothesis of equal forecast accuracy is $H_0: E(d_t)=0 \ \forall \ t$ against the alternative $H_a: E(d_t)\neq 0 \ (\mbox{or} <0 \ \mbox{or} >0)$. Let $\bar{d}$ be the sample mean of the loss differential $\bar{d}=\frac{1}{T}\sum^n_{i=1} d_t$, then by using a HAC estimator for the asymptotic variance of $\bar{d}$; that is $$\widehat{\mbox{Var}(\bar{d})}=\widehat{\gamma}_0 + 2\sum^{K-1}_{k=1} \widehat{\gamma}_k  $$
where  $\widehat{\gamma}_k$ denotes the lag-$k$ sample covariance of the sequence $\lbrace d_t \rbrace $ and $K$ the lag truncation. Under the null hypothesis of equal forecast accuracy we have the statistic,
\begin{eqnarray}
S(1)=\sqrt{T}\dfrac{\bar{d}}{\sqrt{\widehat{\mbox{Var}(\bar{d})}}} \sim N(0,1)
\end{eqnarray}

\section{Data}   
This section describe the variables to be included in all the models and how they should be handle. 
\subsection{Variables Handling}

The series chosen for the Dataset building are some of the main macroeconomics variables that cover the principal determinants of the inflation. The sample of data to be used includes the variables of the main inflation theories that are presented in the literature. Also, in order to measure the different pressures of the different types of goods, we include different subindex of the general prices index. This is where we highlight the different inflationary pressures,

\begin{itemize}
\item Pressure of the prices of goods and services.
\item Pressure of cost.
\item Pressure of exchange.
\item Pressure of expectations.
\item Pressure of money issue and monetary policy rates. 
\item Pressure of demand. 
\end{itemize}
   
The frequency of the data sample are, daily, monthly and quarterly; and the data sample range from January 1st. 2013 until December 31th of 2019. 

The following list describes all the variables in the dataset:

\begin{itemize}
\item {\bf IPCGL: } General Price Index Level of the City of Buenos Aires. Source: GCBA Institute of Statistics. Frequency: Monthly.  
\item {\bf IPCG: } Goods Price Index Level of the City of Buenos Aires. Source: GCBA Institute of Statistics. Frequency: Monthly. 
\item {\bf IPCS: } Services Price Index Level of the City of Buenos Aires. Source: GCBA Institute of Statistics. Frequency: Monthly. 
\item {\bf W: } Mean wage of stable workers (RIPTE). Source: Ministry of Labor, Employment and Social Security of Argentine. Frequency: Monthly.
\item {\bf E:} Nominal exchange peso-dollar. Source: Central Bank of Argentine (BCRA). Frequency: Daily.
\item {\bf E$\pi$:} Inflation expectation. Source: CIF - UTDT. Frequency: Monthly.
\item {\bf Y:} Output, no seasonally Real GDP at constant price. Source: National Institute of Statistics and Census of Argentine (INDEC). Frequency: Quarterly.
\item {\bf M3:} Monetary aggregate. Source:  Central Bank of Argentine (BCRA). Frequency: Daily. 
\item {\bf M:} Money held by the public. Source: Central Bank of Argentine (BCRA). Frequency: Daily.  
\item {\bf r:} Term fixed deposit rate. Source: Central Bank of Argentine (BCRA). Frequency: Daily.
\item {\bf FR:} Financial result, which is defined as Primary result minus debt interest. Source:  Ministry of Treasury of Argentine. Frequency: Monthly.
\item {\bf D:} Total external debt. Source: National Institute of Statistics and Census of Argentine (INDEC). Frequency: Quarterly.   

\end{itemize}

\bigskip

The inflation rate is the growth rate of the General Price Index Level of the City of Buenos Aires (IPCBA) and it will be measured by the logarithmic change of the price index level, similarly we also measure the growth rate of Goods and Services prices,  

\begin{eqnarray}
{\pi_{GL}}_t=\Delta\log(\mbox{IPCGL}_t)=\log(\mbox{IPCGL}_t)-\log(\mbox{IPCGL}_{t-1}),\\
{\pi_G}_t=\Delta \log(\mbox{IPCG}_t)= \log(\mbox{IPCG}_t)-\log(\mbox{IPCG}_{t-1}),\\
{\pi_S}_t=\Delta\log(\mbox{IPCS}_t)=\log(\mbox{IPCS}_t)-\log(\mbox{IPCS}_{t-1})
\end{eqnarray}

\bigskip 

The treated dataset contains time series in different scales and frequencies. Since our study will focus on a monthly base, first we need to transform the data. The process of transformation involved averaging the daily values as monthly values, and the quarterly values were interpolated with a cubic spline procedure as monthly values. Logarithms transform were applied in order to reduce possible source of heteroskedasticity and also to adapt the series to a same scale. Since the Principal Component approach of Stock-Watson requires stationary, the \citet{Dickey and Fuller(1969)} (ADF) test was performed on all series and non-stationary series were differentiated. The lag length for the ADF test were chosen by using the Akaike information criterion (AIC) on a maximum lag order defined by the Schwert rule. Indeed, let $k_{max}=\left[ 12\left(\dfrac{T}{100}\right)^{1/4}\right]$ be the maximum lag suggested by Schwert rule, and let $$AIC(k)=\log(\hat{\sigma}(k))+\dfrac{2k}{T}$$ be the Akaike information criterion for $k$th order ADF regresion model,
$$\Delta y_t = d_t + \rho y_{t-1} + \sum^k_{j=1} \gamma_j \Delta y_{t-j} + u_t, \ (t = k_{max} + 1,...,T), $$           
where $d_t$ denotes the drift or linear trend specification. Since the ADF regression model starts at $k_{max} + 1$, all the competing models with different $k$ are using the same number of effective observations and therefore AIC criterion is comparable selecting the optimal lag as $$k_{opt}=arg \ min_{k\leq k_{max}} AIC(k).$$ 
The ADF test specification was chosen according to the general economic theory and also thought a graphical inspection on all time series; variables regarding to prices, output, exchange and nominal quantities were considered as trending and the variables such as interest rate and expectation were considered as drifting. The results of the ADF test are shown in Table \ref{ADF level test}.

\bigskip

\begin{table}[H]
\centering
{\fontfamily{ptm}\selectfont{
\begin{tabular}{llllll}
\toprule    
 Variable                        & Type         &  Optimal AIC &   $k$     & Test Statistic        & p-value\\ 
\toprule
 $\log$(IPCGL)             & Trend        &  -465.43       &  1      & -1.29     & 0.889    \\
 $\log$(IPCG)           & Trend        & -436.55        &  1      & -0.97     & 0.948          \\
 $\log$(IPCS)        & Trend        &  -421.62       &  2      & -2.35     & 0.403          \\
 $\log$(W)                   & Trend        &  -416.63       &  0      & -2.24      & 0.469          \\
 $\log$(E)                & Trend        &  -218.47       &  1      &-1.82    &   0.697        \\
 $\log$(E$\pi_{GL}$)             & Drift        &  -148.61       &  1      &  -2.15   & 0.017         \\
 $\log$(Y)                  & Trend        & -874.11       &  10      &     -2.37      & 0.397          \\
 $\log$(M3)                      & Trend        &  -411.97       &  8      & -2.36    &  0.398          \\
 $\log$(M)            & Trend        & -367.79       &  11      &   -2.48   & 0.336           \\
 $\log$(r)           & Drift        &  -182.66       &  1      &  -1.50    & 0.068      \\
 $\log$(FR)        & Trend        &   47.83       &  11      &  -1.72    & 0.740     \\   
 $\log$(D)                    & Trend        &  -712.64     &   9        &   -2.32  & 0.422             \\   
\bottomrule                
\end{tabular}
}}
\caption{ADF test at Levels. }\label{ADF level test}
\end{table}

\bigskip
      
The $p$-values used in the test correspond to \citet{MacKinnon(1994)} approximation for the Trend specification. The Drift specification, which  include constant no zero term and no time trend, the Dickey-Fuller test statistic is asymptotically Gaussian due by \citet{Hamilton(1994)} and therefore the classical $p$-values could be used. At the 10 per cent level of significance, the ADF test indicated the presence of unit root in IPCBA price levels, Wages level, Exchange level, Output level, M3 level, Public Money level, Financial result level and for other side the test confirmed the stationary for Interest rate level and Expectation. The non-stationary series were differenced and then ADF test was performed again on all transformed series. Table \ref{ADF diff test} presents the ADF test for all transformed series.

\bigskip          
\begin{table}[H]
\centering
{\fontfamily{ptm}\selectfont{
\begin{tabular}{llllll}
\toprule                              
 Variable                & Type         &  Optimal AIC &   $k$     & Test Statistic        & p-value\\ 
\toprule                         
 $\Delta\log$(IPCGL)   & drift        &     -459.36  &    0      & -4.56     & 9.04e-06       \\
  $\Delta\log$(IPCG)           & drift        &     -431.47  &    0      & -4.36      & 1.9e-05         \\
 $\Delta\log$(IPCS)        & drift        &  -413.98       &  0      & -6.39        & 5.14e-09        \\
  $\Delta\log$(W)                  & drift        &  -408.81       &  0      &   -9.35      & 9.10e-15        \\
 $\Delta\log$(E)               & drift        &   -215.35       &  1      & -6.26   &  9.73e-09        \\
 $\Delta\log$(Y)                  & drift        &    -859.19   &  9      &    -3.45       & 5.067e-04          \\
  $\Delta\log$(M3)                  & drift        &     -403.31         &  7      & -4.47    &  1.6e-05        \\
 $\Delta\log$(M)          & drift        &  -364.51      &  10      &   -2.32      &  .011           \\
  $\Delta\log$(FR)       & drift        &    50.93       &  10      &  -4.82    & 1.3e-05    \\   
 $\Delta\log$(D)                     & drift       &    -683.66          &     6      &  -1.40         &   0.082            \\   
\bottomrule  
\end{tabular}
}}
\caption{ADF test at first differences. }\label{ADF diff test}
\end{table}

\bigskip

After differenced all the above series resulted stationary and therefore integrated of order one. Following this transformation, the $N$ dimensional dataset to be used to form principal component is,

$$X_t = \lbrace {\pi_G}_t,  {\pi_S}_t , \Delta \log(W)_t , \Delta \log(E)_t, $$
$$ \Delta \log(Y)_t , \Delta \log(M3)_t, \Delta \log(M)_t, \log(r)_t, \log(E\pi_{{GL}_t}) , \Delta \log(FR)_t, \Delta \log(D)_t    \rbrace$$

\bigskip

Finally all the series have been standardized to have sample mean zero and sample variance one. 

\section{Factor Analysis}
In this section Factor Analysis is performed by using the Principal Component Methodology, we calculate principal component and also 
calculate the resulting principal components factor estimator $\tilde{F}$. Moreover, we will study the relation between factor and variables.    

We start the analysis by assessing the viability of the dataset $(X_t)$ to achieve Factor Analysis. In this way Kaiser-Meyer-Olkin and Bartlett test  of sphericity were performed on the data sample. Kaiser-Meyer-Olkin is a measure of the sampling adequacy and values greater than 0.5 indicate that the data is acceptable to perform factor analysis and the null hypothesis for Bartlett test is $H_0 : \mbox{variables are not intercorrelated}$. Table \ref{KMO&Bartlett} shows the test performance.

\begin{table}[H]
\centering
{\fontfamily{ptm}\selectfont{
\begin{tabular}{c|c c}

KMO sampling measure        & 0.519                      \\ \hline
Bartlett test of sphericity & Chi-Square         &  219.439        \\
                            & Degrees of freedom &  55              \\
                            & p-value            &  0.000 \\
\end{tabular}

}}
\caption{KMO and Bartlett test on data sample.}\label{KMO&Bartlett}
\end{table}

\begin{figure}[H]  
\begin{center}
\includegraphics[scale=0.65]{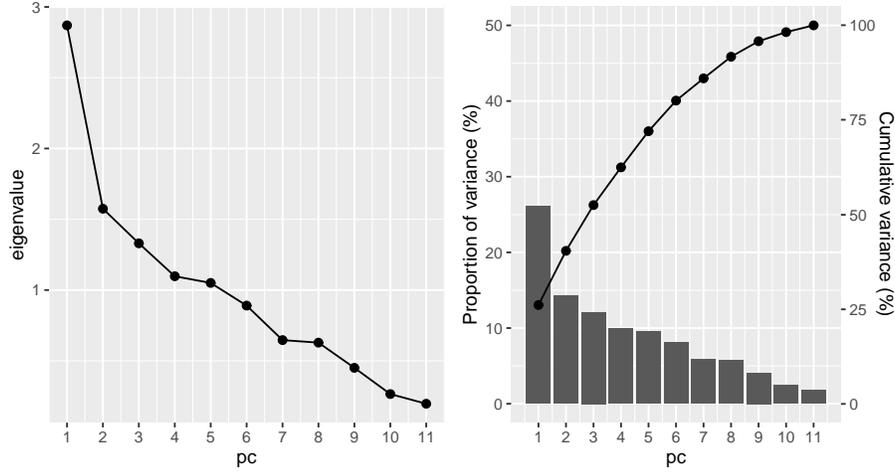} 
\end{center}
\caption{Scree plot and Variance distribution against principal components.}\label{screePlots}
\end{figure}

The tests have shown that the sample dataset $(X_t)$ could be used to perform Factor Analysis. In the next steps we calculate the principal components with the aim of summarize the total variance in the dataset. The eigenvalue greater than one (Kaiser criterion) leads to select five components, however, this is a lower bound for the number of components to extract in principle components
analysis. The scree plot shows that the slope of the graph goes from steep to flat after the forth component and thus suggesting that four factor could be taken. Figure \ref{screePlots} illustrates the scree plot and the variance distribution.

\begin{figure}[H]  
\begin{center}
\includegraphics[scale=0.8]{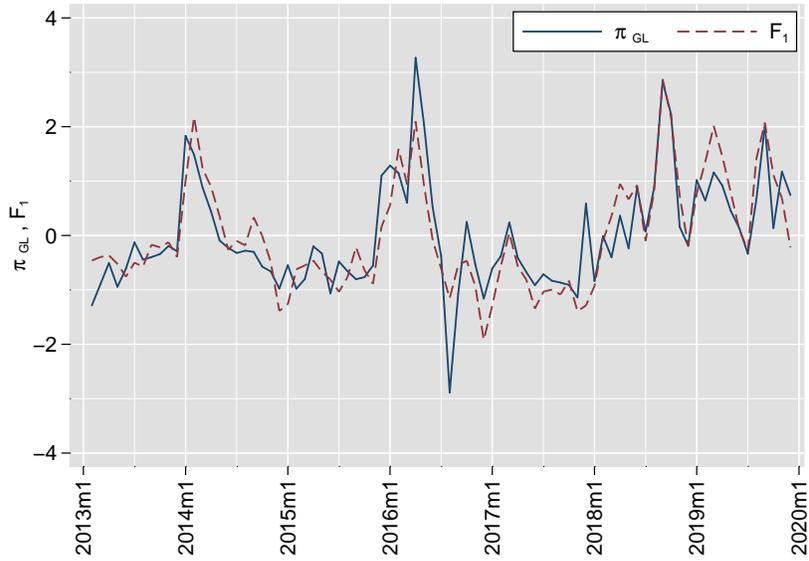} 
\end{center}
\caption{First principal component and standardized monthly inflation.}\label{Componet_vs_Inflation}
\end{figure}

The first four components explain 62\% of the total data variance which is considerable, the first component captures the major share of the variance 26\%, the second 14\%, the third 12\%, and the fourth 9.9\%. Looking at the eigenvectors of the correlation matrix of the sample dataset, it is possible to compute the factors estimators $\tilde{F}$ or scores as the sum of products of the eigenvectors times the values observed for the original variable each period. Equation \ref{primera_componente} and \ref{segunda_componente} describe the estimation of the two main  factors and Figure \ref{Componet_vs_Inflation} illustrate the inflation rate and the constructed factor. 

\begin{eqnarray}\label{primera_componente}
\tilde{{F}_1}_t&=&0.493{\pi_G}_t+0.339{\pi_S}_t+0.233\Delta\log(W)_t + 0.244\Delta\log(E)_t \nonumber \\          
        & +  & 0.380\log(E\pi_{GL})_t +  0.348\log(r)_t -0.285 \Delta \log(M3)_t-0.321\Delta\log(M)_t \\  
        & - & 0.257\Delta\log(Y)_t-0.046\Delta \log(FR)_t - 0.087\Delta \log(D)_t \nonumber \\   
\nonumber
\end{eqnarray}
\begin{eqnarray}\label{segunda_componente}         
\tilde{{F}_2}_t&=&0.234{\pi_G}_t+0.162{\pi_S}_t-0.329\Delta\log(W)_t + 0.481\Delta\log(E)_t  \nonumber\\          
        & -  & 0.080\log(E\pi_{GL})_t + 0.276\log(r)_t + 0.543 \Delta \log(M3)_t + 0.362\Delta\log(M)_t \\  
        & - & 0.043\Delta\log(Y)_t-0.096\Delta \log(FR)_t + 0.242\Delta \log(D)_t \nonumber
\end{eqnarray}

The visual analysis indicates that the first factor show either strong comovements or converse movements with the inflation rate and this should improve predictive abilities.

The correlation between the observed variables and factors may help to understand how the variables are organized in the common factor space and therefore discover clusters of variables. The factor map is a scatter plot between $\mbox{Corr}(X_i,F_1)$ and $\mbox{Corr}(X_i,F_2)$ that shows the relationships between variables and factors; positively correlated variables are grouped together and negatively variables are positioned on opposite sides of the plot origin (opposed quadrants). The quality of representation on the factor map is measured by the squared cosine, for a given variable and a given component we define $ij$-square cosine by,

 $$\mbox{cos2}_{ij}=\mbox{Corr}(X_i,F_j)^2.$$
 
The sum of the $\mbox{cos2}_{ij}$ on all the principal components is equal to one, if a variable is perfectly represented by only two principal components then the sum on these two PCs is equal to one and in this case the variables will be positioned on the circle of correlations. Given a variable we define cos2 as the square norm of the vector $(\mbox{Corr}(X_i,F_1),\mbox{Corr}(X_i,F_2))$ i.e., 

$$\mbox{cos2}= \mbox{Corr}(X_i,F_1)^2 + \mbox{Corr}(X_i,F_2)^2 = \mbox{cos2}_{i1}+\mbox{cos2}_{i2}.$$ 

\begin{figure}[H]  
\begin{center} 
\includegraphics[scale=1.2]{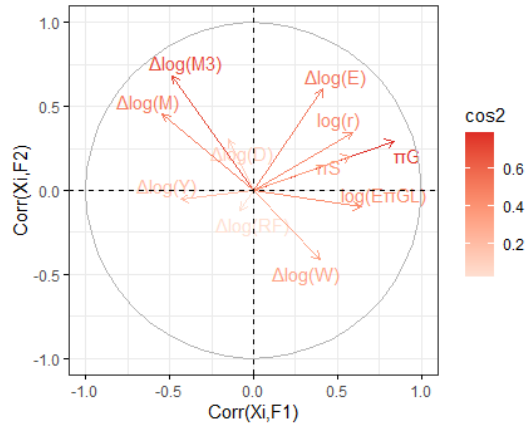} 
\end{center}
\caption{Factor map and cos2 for each variable.}\label{CorrXF}
\end{figure}

Therefore a high cos2 indicates a good representation of the variable on the principal component and in this case the variable is positioned close to the circumference of the correlation circle. On other hand a low cos2 indicates that the variable is not perfectly represented by the PCs and in this case the variable is close to the center of the circle. It’s possible to color variables by their cos2 values and Figure \ref{CorrXF} illustrates the factor map and the cos2 measure by a gradient color map.

In variable study, it is also very useful to look at the variable contributions of each axis for help in interpreting axes.
The contribution of a variable to a given principal component is defined as the ratio between $ij$-square cosine and the total squared cosine of the component; 

\begin{eqnarray}
ctr_{ij}=\dfrac{\mbox{cos2}_{ij}}{\sum_{i} \mbox{cos2}_{ij}}
\end{eqnarray}

\begin{figure}[H]  
\begin{center} 
\includegraphics[scale=1.15]{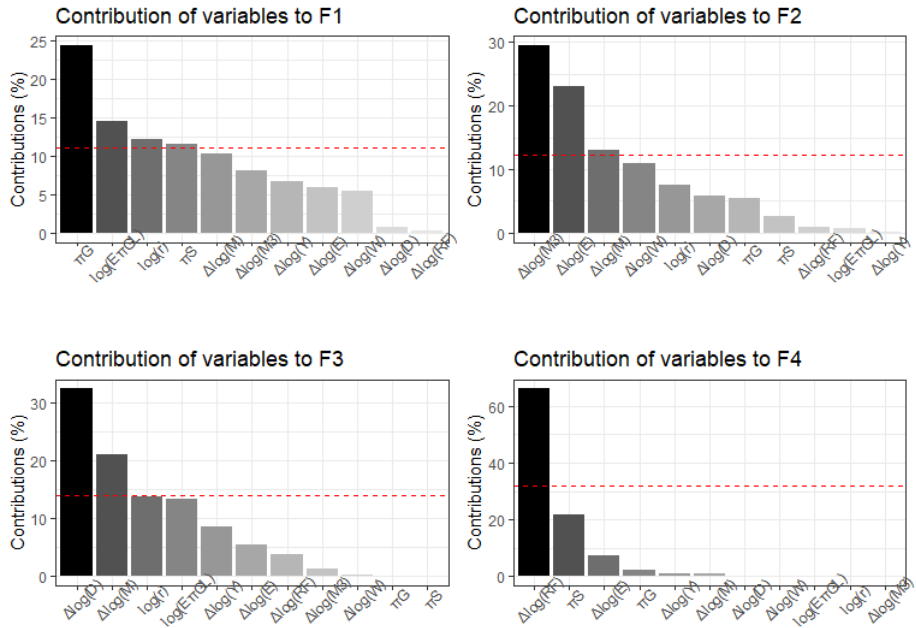} 
\end{center}
\caption{Contribution of the variables to the two principal components.}\label{Var_contribution}
\end{figure}

The larger the value of the contribution, the more the variable contributes to the component. Figure \ref{Var_contribution} illustrates a bar plot of variable contributions for the four principal components,

The red dashed line on the graph above indicates the average of the significantly contributions which could be used as a cutoff to consider a variable as important to the component. Given a factor $j$ the cutoff contribution is calculated as $$\mbox{mean}(crt_{ij}) \ \mbox{subjet} \ \mbox{Corr}(X_i,F_j) \neq 0 \ \mbox{at} \ 10\% \ \mbox{level},$$ then  a variable with a contribution larger than this cutoff could be considered as important in contributing to the component.       
From the contribution graph it can be seen that factors are mainly related to four different groups of variables; the first factor is mainly driven by goods and service prices, inflation expectations and interest rate level. The most important contribution for the first factor is given by goods price with 25\% and then followed by inflation expectations with a contribution close to 15\%. This show that the first factor tends to describe price components. For the second factor, variables such as M3 money growth, exchange rate, money growth by the public contribute the most. In this case the main contribution is driven by M3 money growth with a level close to 30\% and then followed by exchange rate with a contribution level close to 23\%. These finding indicate that the second factor tends to describe money growth and exchanges rate aspects. The third group is mainly related to debt growth with a contribution close to 33\% and the fourth factor is strongly associate with the financial result with a contribution of more than 60\%. On the other hand, Wages and Output have shown a weak contribution into the factors. Our components contributions analysis allowed us to have a better understanding of the economic interpretations of the latent factors.         
To sum up, Table \ref{contribution_groups} shows the groups of variables that contribute the most to the four main factors.

\begin{table}[H]
\centering
{\fontfamily{ptm}\selectfont{
\begin{tabular}{lllll}
\toprule
 $F_1$: Price  & $F_2$: Monetary and Exchange  & $F_3$: Debt   &  $F_4$: Financial  &   \\ 
\toprule                         
   $\pi_G$           &   $\Delta\log(M3)$    &  $\Delta \log(D)$ & $\Delta \log(RF)$ &\\                                                                                       
   $\log(E\pi_{GL})$  &   $\Delta\log(E)$     & $\Delta \log(M) $               &                     &       \\
    $\log(r)$          &  $\Delta\log(M)$   &                  &                     &         \\  
     $\pi_S$           &                     &                 &                     &          \\
                  &           & \\                 
\bottomrule
\end{tabular}
}}
\caption{Main contributions to the components and components interpretations.}\label{contribution_groups}
\end{table}   
 

\section{Factor Model Estimation and Forecast results}
\subsection{Implications for Static Factor regressions }
Consider the regression model given by the main unobserved four factor,

\begin{eqnarray}\label{static_factor_model}
\pi_{GL_t}= \alpha_1 F_{1t} + \alpha_2 F_{2t} + \alpha_3 F_{3t} + \alpha_4 F_{4t} + \varepsilon_t
\end{eqnarray}

\bigskip

Since the estimated factor have economic interpretation we can give economic interpretation to the coefficient when we use as regressor $\tilde{F}_{it}$. Table \ref{results_static_model} shows the estimated model,

\begin{table}[H]\centering
{\fontfamily{ptm}\selectfont{
\def\sym#1{\ifmmode^{#1}\else\(^{#1}\)\fi}
\begin{tabular}{cl}
\hline\hline
Parameters          & Estimation  \\
\hline
$\alpha_1$          &    0.840\sym{***}        \\
                    &   (0.0453)          \\
$\alpha_2$            &  0.297\sym{***}            \\
                    &   (0.0453)             \\
$\alpha_3$           &  -0.012            \\
                    &   (0.0453)             \\
$\alpha_4$            & -0.209\sym{***}            \\
                    &   (0.0453)             \\
\hline
\(R^{2}\)           &     0.839          \\
\hline\hline
\multicolumn{1}{l}{\footnotesize Standard errors in parentheses}\\
\multicolumn{1}{l}{\footnotesize \sym{*} \(p<0.05\), \sym{**} \(p<0.01\), \sym{***} \(p<0.001\)}\\
\end{tabular}
}}
\caption{Factor Model regresion for Equation \ref{static_factor_model} }\label{results_static_model}
\end{table}

The result of the regression showed that factors 1, 2 and 4 were statistically significant at less than 5\% level. According to the interpretation of the factors, the signs obtained are in line with the literature having a positive effect on price level. An increase of one unit in factor one, related to prices, increases the general price level by 0.829\%, while an increase in factor two, associated with monetary growth and the exchange rate, also increases the general price level by around 0.297\%. The effect of the deterioration of the fiscal accounts on the price level is confirmed by the negative sign of factor 4 associated with the financial result, here a drop of one unit generates a 0.209\% increase on prices. On the other hand, factor 3, associated with debt growth, was not significant for any acceptable level of confidence; thus it could suggest that financing of the deficit through debt issuance has a lesser impact on inflation than financing with monetary issue.   

\subsection{Granger-causality and Impulse response functions on FAVAR model}
Hence, we conduct Granger-causality test on the FAVAR model by using the main four factors. The causality tests performed by \citet{Granger(1969)} suggest which variables in the system have significant impacts on the future values of each of the variables in the system. However, the results do not, by construction, indicate how long these impacts will remain effective and impulse response functions may give this information. 
After fitting the FAVAR model we can know whether one variable Granger-causes another, for each equation in FAVAR model we can test the hypotheses that each of the other endogenous variables does not Granger-cause the dependent variable in that equation. We consider the following FAVAR Model by using factors 1,2,3 and 4 with one lag specification, 


\begin{eqnarray}\label{General_FAVAR}
\begin{bmatrix}
\tilde{F}_{1t}\\
\tilde{F}_{2t}\\
\tilde{F}_{3t}\\
\tilde{F}_{4t}\\
{\pi_{GL}}_t
\end{bmatrix}
=\left(
\begin{array}{ccc}
     \phi_{11} & \hdots  & \phi_{15} \\
     \phi_{21} & \hdots  & \phi_{25} \\
     \vdots    & \hdots  & \vdots \\
     \phi_{51}  & \hdots  & \phi_{55}\\
\end{array}
\right)
\begin{bmatrix}
\tilde{F}_{1t-1}\\
\tilde{F}_{2t-1}\\
\tilde{F}_{3t-1}\\
\tilde{F}_{4t-1}\\
{\pi_{GL}}_{t-1}
\end{bmatrix}
+ 
\begin{bmatrix}
\varepsilon_{1t}\\
\varepsilon_{2t}\\
\varepsilon_{3t}\\
\varepsilon_{4t}\\
\varepsilon_{5t}\\
\end{bmatrix}
\end{eqnarray}

\bigskip

Table \ref{Granger} reports Granger tests that the coefficients on all the lags of an endogenous variable are jointly zero. 


\begin{table}[H]
\centering
{\fontfamily{ptm}\selectfont{
\begin{tabular}{clclc}
\toprule
Equation & Excluded & $F$   & df    & $p$-value           \\                
\toprule
               &   $\tilde{F}_2$    &      12.294   &   1     &    0.0008      \\
               &   $\tilde{F}_3$    &      0.237   &   1     &     0.6276         \\
$\tilde{F}_1 $ &   $\tilde{F}_4$    &      0.344   &   1     &     0.5590         \\
               &  $\pi_{GL}$    &   3.938  &     1   &      0.0508       \\
               &  both    &         3.155    &   4  &    0.0187       \\
\midrule
               &  $\tilde{F}_1$    &     2.099    &   1     &    0.1515           \\
               &  $\tilde{F}_3$    &    1.694   &    1   &      0.1969       \\
$\tilde{F}_2 $ &   $\tilde{F}_4$    &     6.165   &   1     &     0.0152         \\               
               &  $\pi_{GL}$    &    1.177   &    1   &     0.2814         \\     
               &  both              &   2.253    &     4    &    0.0711         \\ 
\midrule
               &  $\tilde{F}_1$    &     9.515   &   1     &   0.0028         \\
               &  $\tilde{F}_2$    &     0.0125    &   1     &    0.9111     \\
$\tilde{F}_3 $ &   $\tilde{F}_4$    &      1.923   &   1     &     0.1696         \\               
               & $\pi_{GL}$    &     3.087   &    1   &     0.0829         \\
               &  both              &   4.4831    &     4    &   0.0026        \\ 
\midrule
               &  $\tilde{F}_1$    &    11.478   &   1     &    0.0011         \\
                &  $\tilde{F}_2$    &    4.493    &   1     &    0.0373     \\
 $\tilde{F}_4 $             &   $\tilde{F}_3$    &      8.039    &   1     &    0.0059         \\               
               & $\pi_{GL}$    &     12.388   &    1   &      0.0007          \\
               &  both              &  4.810   &     4    &   0.0016        \\ 
\midrule 
               &  $\tilde{F}_1$    &    6.408   &  1     &     0.0134               \\
                & $\tilde{F}_2$    &   4.287    & 1    &    0.0418        \\
 $\pi_{GL}$    & $\tilde{F}_3$    &  0.198    & 1    &    0.6574        \\
               &   $\tilde{F}_4$    &       2.225   &   1     &     0.1399         \\               
               & both    &   3.826    & 4    &     0.0069      \\
   
\bottomrule                               
\end{tabular}
}}
\caption{Granger causality tests for FAVAR model.}\label{Granger}
\end{table}          

The table \ref{Granger} read as follow, for each equation in the FAVAR system Granger $F$ test is performed by assuming zero the excluded variable. For example, on the first equation for $\tilde{F}_1$ a $F$ test is performed on the lag of $\tilde{F}_2$ and in this case the null hypothesis that $\tilde{F}_2$ does not Granger-cause $\tilde{F}_1 $ is rejected. The second test and third test on the excluded variables $\tilde{F}_3$ and $\tilde{F}_4$ respectively  are not rejected, and the last two test for $\pi_{GL}$ and all excluded variables are rejected. 

In summary we have the following pairwise Granger-Causality maps;
$$\tilde{F}_2 \rightarrow \tilde{F}_1, \ \pi_{GL}\rightarrow \tilde{F}_1, \  \tilde{F}_2 \tilde{F}_3 \tilde{F}_4 \pi_{GL} \rightarrow \tilde{F}_1 $$
$$\tilde{F}_4 \rightarrow \tilde{F}_2, \ \tilde{F}_1 \tilde{F}_3 \tilde{F}_4 \pi_{GL} \rightarrow \tilde{F}_2  $$
$$\tilde{F}_1 \rightarrow \tilde{F}_3, \  \pi_{GL}\rightarrow \tilde{F}_3, \  \tilde{F}_1 \tilde{F}_2 \tilde{F}_4 \pi_{GL} \rightarrow \tilde{F}_3 $$
$$ \tilde{F}_1 \rightarrow \tilde{F}_4,\ \tilde{F}_2 \rightarrow \tilde{F}_4, \ \tilde{F}_3 \rightarrow \tilde{F}_4, \ \pi_{GL} \rightarrow \tilde{F}_4, \ \tilde{F}_1 \tilde{F}_2 \tilde{F}_3 \pi_{GL} \rightarrow \tilde{F}_4 $$
$$ \tilde{F}_1\rightarrow \pi_{GL}, \ \tilde{F}_2\rightarrow \pi_{GL}, \ \tilde{F}_1 \tilde{F}_2 \tilde{F}_3 \tilde{F}_4 \rightarrow \pi_{GL} $$

According to the factors identifications, these results are in line with the literature. Since the second factor is associated to unobserved facts of Monetary and Exchanges rates aspect it is highly probable that this variable may have a contemporaneous effect on the first factor which is represented as inflation expectation and goods price and therefore effect on the general level of the inflation rate. Another economic interpretation is given with factor 4 and factor 2, responding to the financing budget equation.

\begin{figure}[H]  
\begin{center}
\includegraphics[scale=0.9]{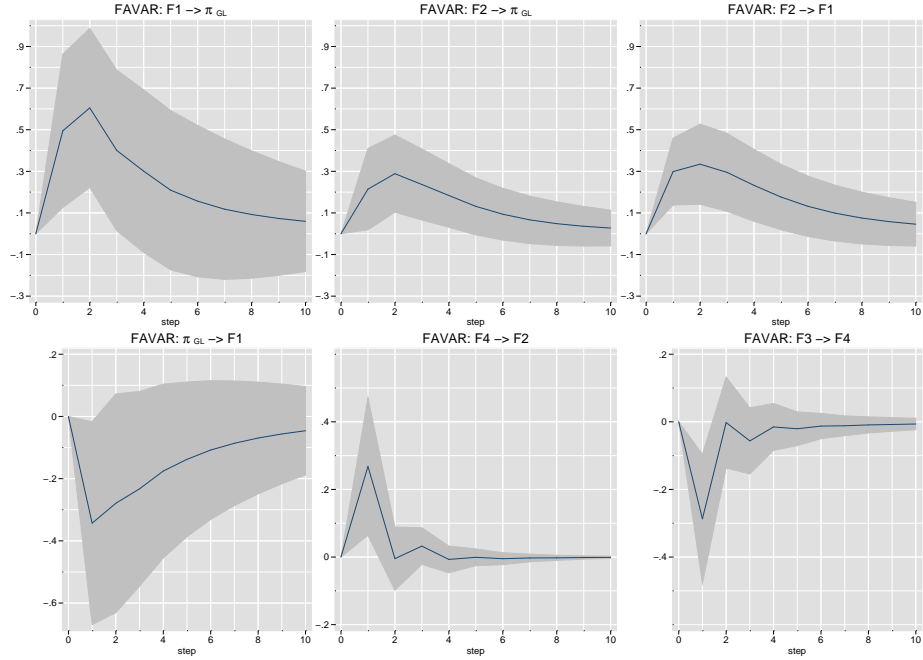} 
\end{center}
\caption{Impulse Response Function for FAVAR model.}\label{Impulse-response}
\end{figure}

The Impulse Response Functions trace out responsiveness of dependent variables in the VAR to shocks to each of the
variables. A distinguishing feature of these generalized approaches is that the results from these analyses are invariant to the
ordering of the variables entering the VAR system. For each variable from each equation separately, a unit shock is applied to the error, and the effects upon the VAR system over time are noted. The results of the impulse response functions are presented in Figure \ref{Impulse-response}.

According to these results, the general level inflation rate response from a unit shock in the first factor has a positive response and after one month it was close to 0.6. The conventional interpretation for this is that shocks to the first factor are mainly driven by inflation expectations and goods prices and therefore responding to inflation pressures. Shocks to the second factor has also a positive effect on inflation rate with a maximum level of 0.3 after two months. Since the second factor is mostly driven by monetary and exchange facts, the inflation response is in line with what is expected. The first factor response from a shock in the second factor is also expected; the exchange rate contribution of the second factor impulse tradable goods and therefore inflation expectations. Response on factor two from a shock on factor 4 is also expected, since factor 4 is related to financial result an improvement of it should be in line with an increasing of the monetary issue and therefore on the aggregate. The fourth factor response from a shock on factor 3 cuold be explained by the increasing of the participation of the interest rate for debt into the financial result balance.

\subsection{Forecast} 
In this section we perform the Factor Model following the methodology given in Section \ref{theoFactormodel}. The aim is to forecast the inflation rate $\pi_{GL}$ under the two forecasting approach by using out of the sample data on different forecast horizon. The benchmark model chosen to compare was an autoregressive model (AR) whose structure was determined through the analysis of the partial autocorrelation function and the Akaike information criterion and the resulted order was one. The model performance is compared against an AR(1) benchmark model and the relative RMSE and U-Theil measure are calculated. Low values of RMSE and U-Theil indicate smaller forecast error and the lowest relative RMSE and U-Theil indicates the highest predictive abilities. Also \citet{Diebold and Mariano(1995)} (DM) test is performed against benchmark by using the Stata module created by \citet{Baum(2003)}, which follows the theoretical framework given in Section \ref{DM}. The loss function considered is $g(e_{it})=e_{it}^2$ and the autocovariance of the loss differential is given by taking the lag structure close to the cubic root over the number of observations. Since the test is a two-sided, the rejection of the null hypothesis against the alternative suggest that if the S(1)-statistic is negative the benchmark model is preferred, and if S(1)-statistic is positive Factor Model is preferred. Three different forecast horizon are defined; the first ($P_1$) from January 2019 until December 2019, the second ($P_2$) from May 2019 until December 2019 and the third ($P_3$) from September 2019 until December 2019. 

\subsubsection{Performance for Static forecast}

Here we present the performance for the forecast model given by Equation \ref{static_forecast} by using one factor and two factors. Specifically, we consider the following two Factor Model forms (1 FM) and (2 FM) respectively,

\begin{eqnarray}
{\pi_{GL}}_t&=& \eta+{\pi_{GL}}_{t-1}+\alpha_1\tilde{F_1}_{t-1}+\varepsilon_{t} \nonumber \\
{\pi_{GL}}_t&=&\eta+{\pi_{GL}}_{t-1}+\alpha_1\tilde{F_1}_{t-1}+\alpha_2\tilde{F_2}_{t-1}+\varepsilon_{t} \nonumber
\end{eqnarray}

\begin{figure}[H]  
\begin{center}
\includegraphics[scale=1]{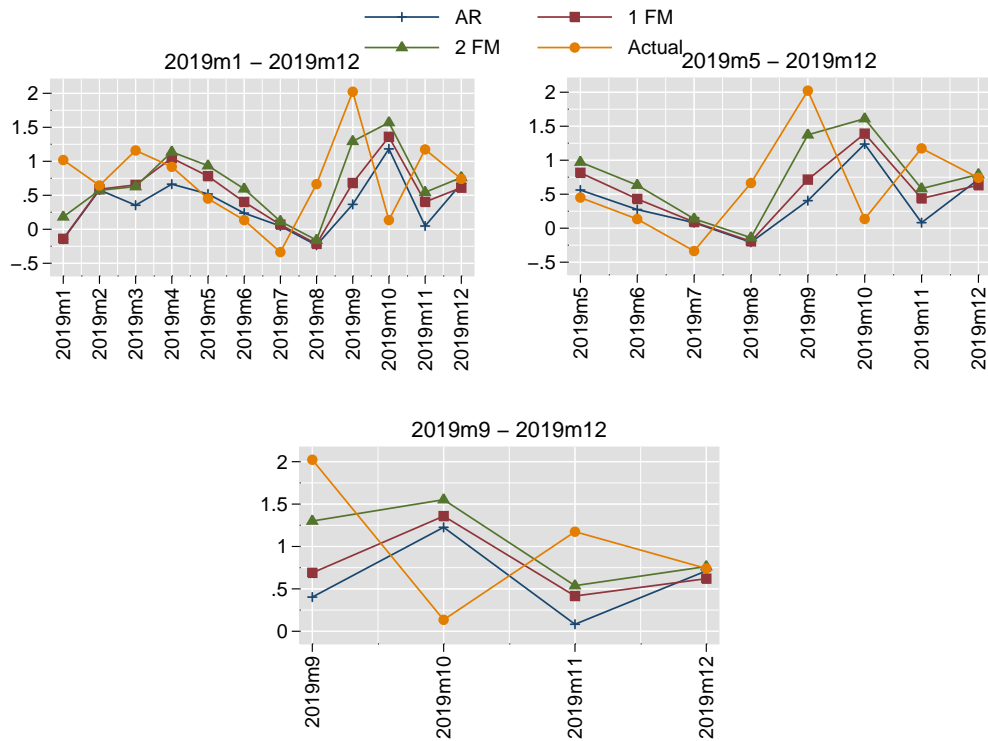} 
\end{center}
\caption{Comparison of the Static inflation forecasts using different models and different
out of the sample horizon.}\label{forecastEstatico}
\end{figure}

Figure \ref{forecastEstatico} illustrates the static forecasting and Table \ref{RMSEperformance} presents the forecast performance under RMSE, U-Theil measures and Diebold and Mariano test results.  
              
\begin{table}[H]
\centering
{\fontfamily{ptm}\selectfont{
\begin{tabular}{*9c}
\toprule
Period & Model & $RMSE_P$   & $U_P$    & $RRMSE_P$   & $RU_P$     & \multicolumn{3}{c}{DM-test}        \\                
\toprule
       &        &            &             &         &            & S(1)-stadistic   & $p$-value    &   $K$ lag       \\
       &  AR    &   0.824    &   0.565     &  1      &  1         &                  &            &         \\
$P_1 $ & 1 FM    &   0.744    &  0.461   &    0.902     &  0.815  &     2.790         & 0.0053 &    3        \\
       &  2 FM    &  0.667    &  0.378    &   0.809     & 0.669   &     2.647         & 0.0081 &    3        \\
\midrule
       &  AR    &   0.865    &  0.577     &    1     &  1         &                  &            &             \\
$P_2$  & 1 FM    &  0.788     & 0.485    &     0.910     &  0.840 &     1.660          &  0.0969  & 3        \\
       & 2 FM    &  0.736     & 0.398    &    0.850     &   0.689 &     1.961        &  0.0499    & 3         \\
\midrule
       &  AR    &     1.118   & 0.568     &    1      &  1        &                  &               &      \\
$P_3$  & 1 FM    &   0.982    &  0.473    &    0.878     &  0.832 &    1.974        &   0.0483    & 2     \\
       & 2 FM    &    0.856    & 0.365    &    0.765     & 0.642  &     1.618     &  0.105        & 2          \\
    
\bottomrule                               
\end{tabular}
}}
\caption{Static Forecast evaluations criterions. }\label{RMSEperformance}
\end{table}          

All ratios of the Root Mean Square Error and U-Theil against autoregresive model are less than unity on all horizon and also it can be seem that the 2 Factor Model is better than the 1 Factor Model on all horizon. The S(1)-statistic was statistically significant at the 10\% level and therefore suggesting that factors extracted from observed group of series have a certain potential in forecasting the dynamics of consumer prices index.

\subsubsection{Performance for Dynamic forecast}
In order to perform the Dynamic forecasting Equation \ref{dynamic_forecast1} was performed to get the forecast of factor estimator and then the inflation forecast was computed by using Equation \ref{dynamic_forecast2}. Specifically, we consider the following two FAVAR Model with one lag specification, 

\begin{eqnarray}
\mbox{FAVAR1} : 
\begin{bmatrix}
\tilde{F}_{1t}\\
{\pi_{GL}}_t
\end{bmatrix}
=\left(
\begin{array}{cc}
     \phi_{11} & \phi_{12}  \\
     \phi_{21} & \phi_{22}  \\
\end{array}
\right)
\begin{bmatrix}
\tilde{F}_{1t-1}\\
{\pi_{GL}}_{t-1}
\end{bmatrix}
+ 
\begin{bmatrix}
\varepsilon_{1t}\\
\varepsilon_{2t}\\
\end{bmatrix}
\end{eqnarray}

\begin{eqnarray}\label{General_FAVAR}
\mbox{FAVAR2} : 
\begin{bmatrix}
\tilde{F}_{1t}\\
\tilde{F}_{2t}\\
{\pi_{GL}}_t
\end{bmatrix}
=\left(
\begin{array}{ccc}
     \phi_{11} &  \phi_{13} & \phi_{13} \\
     \phi_{21} & \phi_{22}  & \phi_{23} \\
     \phi_{31}  & \phi_{32}  & \phi_{33}\\
\end{array}
\right)
\begin{bmatrix}
\tilde{F}_{1t-1}\\
\tilde{F}_{2t-1}\\
{\pi_{GL}}_{t-1}
\end{bmatrix}
+ 
\begin{bmatrix}
\varepsilon_{1t}\\
\varepsilon_{2t}\\
\varepsilon_{3t}\\
\end{bmatrix}
\end{eqnarray}

\begin{figure}[H]  
\begin{center}
\includegraphics[scale=1]{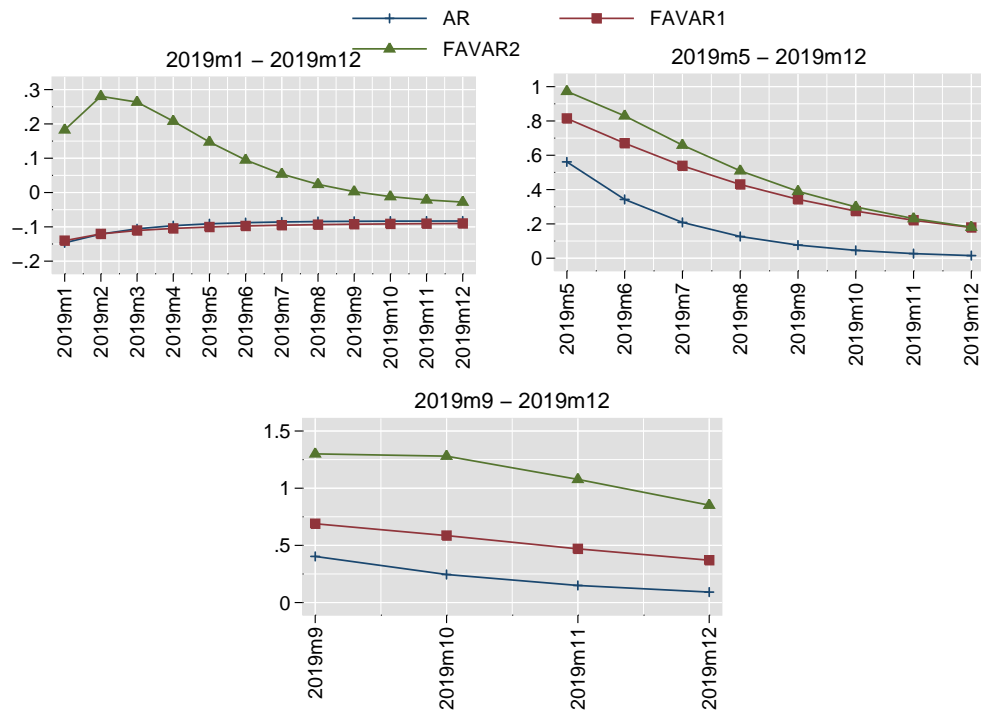} 
\end{center}
\caption{Comparison of Dynamic inflation forecasts using different models and different
out of the sample horizon.}\label{forecastPicture_dy}
\end{figure}

        
Figure \ref{forecastPicture_dy} illustrates the dynamic forecasting and Table \ref{RMSEperformance_dy} presents the forecast performance under RMSE and U-Theil measures. 
\begin{table}[H]
\centering
{\fontfamily{ptm}\selectfont{
\begin{tabular}{*9c}
\toprule
Period & Model & $RMSE_P$   & $U_P$    & $RRMSE_P$   & $RU_P$      & \multicolumn{3}{c}{DM-test}           \\                
\toprule                                                            
     &        &            &             &         &               & S(1)-stadistic   & $p$-value       & $K$ lag        \\
       &  AR    &   1.011    &   0.981     &  1      &  1          &                  &               &          \\
$P_1 $ & FAVAR1    &   1.016    &  0.980   &    1.004     &  0.998 &        -3.370      &    0.0008& 3         \\
       & FAVAR2   &   0.862    &  0.797    &   0.852     & 0.812  &          2.311      &   0.0208 & 3            \\
\midrule
       &  AR    &   0.886    &  0.755     &    1     &  1          &                  &               &               \\
$P_2$  & FAVAR1    &  0.814     & 0.579    &     0.918     &  0.766&       0.9217     &   0.3567   &  3             \\
       & FAVAR2  &  0.841     &  0.561    &    0.949     &   0.743 &       0.4282     &   0.6685   &  3  \\
\midrule
       &  AR    &     1.013   & 0.684     &    1      &  1         &                  &               &     \\
$P_3$  &  FAVAR1    &  0.808     &  0.456    &    0.797     &  0.666  &    5.155   &  0.0000       &  2   \\
       &  FAVAR2    &  0.681    & 0.287    &    0.672     & 0.419     &    3.276   &  0.0011       &  2    \\
    
\bottomrule                               
\end{tabular}
}}
\caption{Dynamic Forecast evaluations criterions. }\label{RMSEperformance_dy}
\end{table}  
 
For dynamic forecast the best performance was achieved for FAVAR2 on the first and third period showing both ratios far less than the unity and also with the S(1)-statistic statistically significant at the 10\% level. For the second period FAVAR1 and FAVAR2 performed very similar for both metrics and also better than benchmark in terms of the performance measures, however S(1)-statistic did not result statistically significant in both cases.

\section{Conclusion}
This paper has two main goals; firstly investigates the relation between several macroeconomics variables and the inflation rate by using Factor Analysis and secondly focuses on the performance of Factor Model when it is used to predict the Argentinian inflation rate. The study of contributions indicated that the four main factors have an economic interpretation. The first factor is strongly associated with the aspects of goods prices and inflation expectations, the second one deals with monetary growth and exchange aspects, the third is related to debt growth and the fourth is related to financial results. The identification of the static factor model allowed us to interpret the ceteris paribus effect of each factor on the inflation rate. Here factors 1 and 2 had positive effects, with factor 1 having the greatest impact. The effect of factor 3 was close to zero and statistically not significant, evidencing a possible low impact of debt growth on inflation. On the other hand, factor 4 was negative, thus positively impacting the inflation rate against the deterioration of the financial result. Output and wages have not shown significant contributions on the main factors and therefore evidencing that inflation is far to be a problem related with aggregate demand and wages. Granger-Causality tests and the impulse response functions also suggested a positive effect of factors 1 and 2 on inflation and therefore identifying the crucial role of goods prices, inflation expectation, monetary growth and exchange in driving the dynamics of inflation. Secondly, we have presented forecasting models by using one and two factors and the findings provided evidence that factor model forecasts improve upon the benchmark model on static and dynamic forecasting. Thus, factors extracted from observed group of series have a certain potential in forecasting the dynamics of consumer prices index. 
All of these findings indicate that multivariate analysis allows a better understanding of the sources of the different inflation drivers.  
%
%
%
%

\end{document}